# An Efficient Distributed Data Extraction Method for Mining Sensor Network's Data

Azhar Mahmood[1], Ke Shi[1] and Shaheen Khatoon[1]

[1] School of Computer and Applied Technology
Huazhong University of Science & Technology (HUST) Wuhan, China

**Abstract**
A wide range of Sensor Networks (SNs) are deployed in real world applications which generate large amount of raw sensory data. Data mining technique to extract useful knowledge from these applications is an emerging research area due to its crucial importance but still it's a challenge to discover knowledge efficiently from the sensor network data. In this paper we proposed a Distributed Data Extraction (DDE) method to extract data from sensor networks by applying rules based clustering and association rule mining techniques. A significant amount of sensor readings sent from the sensors to the data processing point(s) may be lost or corrupted. DDE is also estimating these missing values from available sensor reading instead of requesting the sensor node to resend lost reading. DDE also apply data reduction which is able to reduce the data size while transmitting to sink. Results show our proposed approach exhibits the maximum data accuracy and efficient data extraction in term of the entire network's energy consumption.

## 1. Introduction

Advances in wireless communication and microelectronic devices led to the development of low power sensors and the deployment of large scale sensor networks. With the capabilities of pervasive surveillance sensor networks has attracted significant attention in many applications domains, such as habitat monitoring [1, 2], object tracking [3, 4], environment monitoring [5-7], military [8, 9], disaster management [10], just to mention a few example[11]. These applications yield huge volume of dynamic, geographically distributed and heterogeneous data. The raw data if analyzed in an appropriate way might help to automatically and intelligently solve a variety of tasks thus making the human life more safe and comfortable. Recently, extracting knowledge from sensor data has been received a great deal of attention by the data mining community. However, the extremely constrained nature of sensors and the potentially dynamic behavior of SNs hinder the use of traditional mining approaches commonly applied on other domains. Traditional approaches are meant for multi-step methodologies and multi-scan algorithms, which cannot be straightforwardly applied to sensor network. Development of algorithms that consider the characteristics of sensor networks, such as energy and computation constraints, network dynamics, faults, constitute an active area of current research.

Several techniques have been proposed in the literature for knowledge extraction from sensor data e.g. association rules [12-14] frequent patterns mining, knowledge discovery over data streams[15, 16], and clustering [17] to enhance the performance of SNs. In these applications large numbers of sensors are distributed in the physical world and generate streams of data that need to be combined, monitored, and analyzed on central side. However, collecting all data in a central computing node with a high computational power does not optimize the use of energy-costly transmissions. Indeed in most cases all raw data are not needed, we are only interested in an estimate of a small number of parameters. Instead of computing such parameters on the sink node, a better approach suggests that each node contributes to the computation. Since accessing the data, processing data, and transmitting data are all tasks that consume energy which is a limited resource in sensor node. So, what should be the solution for theoretical and applicative research in SNs for efficient data extraction? This question motivates us to develop a distributed data extraction (DDE) method which pre-processes the raw data directly at sensor node. Hence, instead of sending the raw data to the central site, sensor nodes use their processing abilities to locally carry out simple computations and transmit only the required and pre-processed data. The processing performs at each sensor node is helpful for taking real time decision as well as can serve as prerequisite for development of scalable data mining technique on central side. In DDE method the major contributions are following:

1. Rule based clustering technique for efficiently extracting data from sensors nodes to optimize network lifetime in term of energy and data size. These rules are identified by applying association rule mining on cluster head (CH) node.
2. A significant amount of sensor readings sent from the sensors to the data processing point(s) may be lost or corrupted. In DDE this problem is addressed by estimating missing values from available sensor reading instead of requesting the sensor node to resend lost reading. The key advantage of our missing value estimation is that it is done directly at sensor node and can be used to identify the behavior of the sensor nodes. Data Reduction is applied which is able to reduce the data size received from sensor nodes. The extracted data is more compact than raw sensor data and can therefore be more efficiently transmitted to sink from the sensor network.

The rest of this paper is organized as follow: after introducing basic concept of SNs data mining in Section 1; we provided an overview of related work of data extraction methods either centralized or distributed in Section 2. Proposed method, algorithms and its details are presented in section 3; Simulation results are presented in section 4 and finally sections 5 concludes the paper and suggest directions for the future work.

## 2. Related Work

Several techniques have been proposed in the literature to enhance the performance of SNs, such as frequent pattern mining, clustering, classification, prediction, just to mention a few examples. In this section we review past studies in term of three categories related to this research: Association Rule Mining, Missing value identification and Clustering methods.

Tanbeer et al. [18] and Boukerche and Samarah [12] proposed centralized data mining models to find association among the sensors nodes. They proposed tree-based data structure that used FP-growth approach to obtain the frequency of all events detecting sensor. Tanbeer et al. used Sensor Pattern Tree (SP-Tree) to construct a prefix-tree and reorganize the tree in a frequency descending order. Through the reorganization the SP-tree can maintain the frequently event-detecting sensors' nodes at the upper part of the tree, which provides high compactness in the tree structure. Once the SP-tree is constructed FP-growth mining technique is applied to find the frequent event-detecting sensor sets. Boukerche, and S. Samarah [19] used Positional Lexicographic Tree (PLT) structure for mining association rules in which the event-detecting sensors are the main objects of the rules regardless of their values. The mining begins with the sensor having the maximum rank by generating the frequent patterns from its PLT in a recursive way. The computation required at each recursion to update the PLT involved in the prefix part of a pattern. Therefore, the two database scans requirement and the additional PLT update operations during mining limit the efficient use of this approach in handling SNs data. K Romer, [20] and Chong et al. [21] link the problem of mining sensor data to the association rules' mining problem by proposing in-network models. Romer's approach takes into consideration the distributed nature of wireless sensor networks to discover frequent patterns of events with certain spatial and temporal properties. Whereas, Chong et al. finds strong rules from sensor readings and use these learnt rules as a triggers to control sensor network operations or supplement sensor operations. For example, triggers activated from the rules could be used to sleep sensors or reduce data transmissions to conserve sensor energy. Our proposed in-network technique is different from Romer's and Chong et al. approach in a way that extracted rules are used to cluster the sensor node and estimating missing sensor's values.

For missing values identification Halatchev and Gruenwald [22] proposed a centralized methodology called Data Stream Association Rule Mining (DSARM) to identify the missing sensor's readings. It uses Association Rule Mining algorithm to identify sensors that report the same data for a number of times in a sliding window called related sensors and then estimates the missing data from a sensor by using the data reported by its related sensors.

For the clustering issues in sensor networks, several methods have been proposed. Clustering protocol for node clustering such as LEACH [23], ACE [24], HEED [25], DEEH [26] and Energy Aware Protocol (EAP) [27] are proposed to solve energy consumption problems in SNs. These protocols probabilistically selects several nodes as cluster heads according to their residual energy, and then remainder nodes are joined into clusters to minimize the communication cost between them and corresponding cluster heads. Yoon and Shahabi [28], Beyens et al. [29], Yeo et al. [30] proposed data correlation clustering architecture for WSNs in which cluster-heads spatiotemporally correlate. In Beyens et al. approach cluster head maintains a local prediction model that is used to select a suitable node of the cluster to be activated. The idea is to put a sensor node to sleep when there are no objects in its sensing region. In Yoon and Shahabi approach nodes are groups based on similar values and only one reading per group is transmitted. Whereas, in Yeo et al. approach the size of data size is reduced at each cluster head by applying data suppression technique.

All the above techniques have focused on extracting data regarding the phenomenon monitored by the sensor nodes, in which the mining techniques are applied to the sensed data received from the sensor nodes and accumulated at a central database. In our work, we have proposed an in-network data extraction approach to extract the pre-processed sensor data required for mining by applying rule based clustering to save energy and in-network missing value estimation to increase the accuracy of extracted data. Furthermore a data reduction method is used to reduce the transmission energy and data size.

# 3. Proposed Distributed Data Extraction (DDE) Method

In this section we proposed distributed data extraction methodology for efficiently extracting data from SNs. The main goal is to overcome the challenges for mining continuous stream of data arrived from SNs. We adopted distributed solution where sensor nodes are using their processing capabilities to perform computation and instead of sending the raw data, preprocessed data should be transmitted from nodes to sink. The system workflow consists of three main phases: (1) Clustering of sensor nodes (2) identification of missing sensor and estimation of value (3) data reduction. Our clustering and missing value identification methods are based on association rule mining. To apply association rule mining in SNs we first define association rule mining problem for sensor network.

## 3.1 Association Rules Mining Problem in Sensors

The association rule mining problem define for transactional database are develop to work on static data and cannot be applied directly on SNs data, where the data is continuous and come with high speed. Static data base algorithms require multiple scans of the original database, which leads to high CPU and I/O costs. Therefore, they are not suitable for a SNs data, in which data can be scanned only once. In view of these challenges we aim to define sensor association rule mining problem. The definition of mining sensor association rules use in our DDE approach following the definition provided by Boukerche and Samarah[19] inspired by the definition of frequent patterns proposed in domain of transactional database by Agrawal et al. [31].

Let $S = \{s_1, s_2, \ldots s_n\}$ a set of sensors in a particular sensor network. We assume that the time is divided into equal-sized slots $(t_1, t_2, \ldots t_w)$ such that $t_{w+1} - t_w = \lambda$ for all $1 < w < n$, where $\lambda$ is the size of each timeslot, and $T_{his} = t_n - t_1$ represents the historical period of data during data extraction process. The main step in the formation of association rules is to find the patterns of sensors that co-occur together and exceed a certain frequency (these patterns are called frequent patterns). After finding the frequent patterns association rules are generated. For instance, the rule $(s_1 s_2 \rightarrow s_3)$ is generated from the pattern $(s_1 s_2 s_3)$.

**Definition 1.1.** Suppose sensors data is stored in epoch, where each epoch contains time slot, sensor id and sensor value which sense in given time slot. Let $P = \{s_1, s_2, \ldots s_k\}$ is a set of sensors that detect event in same time slot ($Dts$) and node value $N_V = \{v(s_1, s_2, \ldots s_k)\}$ then an epoch $D$ is defined as following: $D(Dts, P, N_V)$.

Given a database of epochs (DS) generated after a particular historical period, the problem of mining sensors' association rules is to generate all the rules present in the DS.

**Definition 1.2.** The frequency freq of the pattern P in DS is defined to be the number of epochs in DS that supports it.

**Definition 1.3.** Let min sup represent the minimum number of epoch that P should satisfy. The P is said to frequent if its freq is greater than the min sup i.e.
$Freq(P, DS) = \{D(D_{ts}, P\} \geq min\ sup$.

**Definition 1.4.** Sensor association rules between two sensor $s_1$ and $s_2$ in P are implication of form $s_1 \rightarrow s_2$ where $s_1, s_2 \subset S$ and $s_1 \cap s_2 = \phi$

**Definition 1.5.** Support and confidence of the rule $s_1 \rightarrow s_2$ is defined as follow:
$Freq(s_1 \rightarrow s_2) = (s_1 \cup s_2, DS)$
$Conf(s_1 \rightarrow s_2) = freq(s_1 \cup s_2, DS) / freq(s_1)$

The rule $(s_1 \rightarrow s_2: 90\%, \lambda)$ means if we receive events from sensors $s_1$ then there is a 90 % chance of receiving an event from sensor $s_2$ within $\lambda$ units of time. Note that frequency and support are used interchangeably and min *sup* represents the minimum number of epochs that the frequency of the rules should satisfy. The main challenges of mining these rules can be as follow:

1. How data can be extracted efficiently from the sensor network required for mining process
2. How the patterns that meet the given minimum support can be generated efficiently

## 3.2 Data Extraction Methodology

The network architecture used for extracting the data is shown in Fig.1. It consists of sensor nodes deployed randomly and the network is divided into groups based on distance from the sink. Each group has its own cluster number and member nodes. The database is attached at sink to store the preprocessed data from each Cluster-Head (CH).

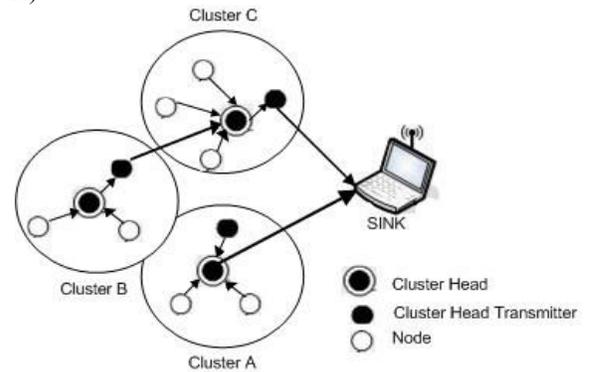

Fig.1. Network Model

1. N sensors are randomly deployed within circular field *A*. The sink is deployed far away from *A*

2. Every node and sink is at fixed position; the location of sink and distance is known to each node and can communicate directly to sink
3. CH nodes uses clustered based multiple-hop mode of transmission to route the data towards sink
4. All nodes are homogenous means each have same capacities

Table.1 Notations used in algorithms

| Notation | Meaning | Notation | Meaning |
|---|---|---|---|
| $H_P$ | Historical Period | $T_S$ | Timeslot |
| $S$ | Support | $N_L$ | Node Location |
| $S_L$ | Sink Location | $N_V$ | Node Value |
| $R_L$ | Rules | $N_{TE}$ | Node Total Energy |
| $R_N$ | Range | $N_{ID}$ | Node ID |
| $C_D$ | Cluster Distance | $CH_{ID}$ | Cluster Head ID |
| $C_I$ | Confidence | $CHT_{ID}$ | Cluster Head Transmitter ID |
| $C_H$ | Cluster Head | $CHT$ | Cluster Head Transmitter |
| $N_F$ | Node Frequency | | |

The data extraction process is shown in Algorithm 1. The notations used in algorithms are shown in Table 1. Algorithm 1 shows the data extraction process starts with the application that provides the mining parameters to the sink which includes *Timeslot size $T_S$, Historical Period $H_P$, Support $S$, Range $R_N$, Cluster Distance $C_D$, Rules $R_L$* and *Confidence $C_I$*. The Sink broadcast these parameters to the network nodes. The nodes collect data and transfer it from *node→CH→Sink* or *node→CH→CHT→Sink* in multi-hop fashion. In this way computation load is distributed on sensor nodes especially on CH nodes within network.

### 3.2.1 Cluster Formation

Algorithm 2 shows the cluster formation process. At the end of each $T_S$ network nodes checks its sensed data and broadcast messages to nodes within given cluster distance $C_D$ for cluster formation. Cluster formation uses the $R_N$ and $C_D$ to group the sensor in same cluster. Upon receiving the broadcasted message each node checks the value of $R_N$. If its value is within $R_N$ it saves in its buffer and compares $C_D$ with each node's distance. If the distance between nodes is less than or equal to $C_D$ and sensed value is within given $R_N$ then those group of nodes forms a cluster.

In the second round association rules are scanned first for cluster formation. The nodes $N_{ID}$ which are associated they will not broadcast message for cluster formation. These nodes join same cluster within $C_D$. Nodes which are not will join cluster formation process based on $R_N$ and $C_D$. e.g. if rules says $S_1S_2 \rightarrow S_3$, $S_1S_3 \rightarrow S_2$, $S_2S_3 \rightarrow S_1$, in this case $S_1$, $S_2$, $S_3$ nodes are in same cluster and only participate in cluster head selection step. These nodes will not participate in cluster formation process in upcoming rounds which save sensor's energy and reduced number of messages broadcast.

*Algorithm1. DDE*
*Input: Raw Data Stream (DS)*
*Output: Pre-Processed Data (PS)*
*SINK:*
*Broadcast parameters($H_P$, $T_S$, $S$, $R_N$, $R_L$, $C_D$, $C_I$)*
*Upon Receiving all messages*
*For Slot Number=1to($H_P/T_S$)*
*P=The set of the sensors identifies with in the same timeslot*
*D=(Slot Number, P)*
*Insert(D,DS)*
*NODE:*
*SET CHFound=False*
*TimeSlot=1*
    *For (i=1; to $H_P/T_S$; i++)*
    *Sense Data($N_{ID}$, $T_S$);*
    *Broadcast ($N_{ID}$, $T_S$, $T_E$, $N_L$, $N_V$)*
    *For (Network Nodes i to n)*
    *ScanRules ($R_L$)*
    *If (ScanRules ($R_L$)==False)*
    *{*
    *ClusterFormation ($N_{ID}$, $T_S$, $T_E$, $N_L$ )*
    *{*
    *Range Datagroup( 1 to n) within $C_D$ and $R_N$*
    *MatchRulesID($R_L$)*
    *CalculateDiatance($N_{ID}$, $N_L$)*
    *Return $CH_{ID}$, $CHT_{ID}$*
    *}}*
*Else*
*Join($N_{ID}$ Clsuter) //If ScanRules() return True then join cluster within given $C_D$*
*SET CHFOUND=True*
*CHBroadcast ($CH_{ID}$)*
*CHEpoch=TransferData($N_{ID}$, $N_v$)*
 *TimeSlot= TimeSlot++*
*}*
*MissingValues(SensorAssociationEpoch);*
*DataEstimation(CHEpoch)*
*ApplyReducation(PEpoch)*
*//@TansfeEnergy=Amount of Energy required to transmit Epoch to Sink*
*If($CH_{ID}$ $T_E$<@TransferEnergy)*
*{*
*Transmit data $CHT_{ID}$ (PreprocessedEpoc)*
*Send to SINK (PreprocessedEpoch DS, Rules $R_L$)*
*}*
*Else*
*Send to SINK (PreprocessedEpoch DS, Rules $R_L$)*

### 3.2.2 Cluster Head Selection

Upon completion of cluster formation process each node has its cluster members $N_{ID}$, node location $N_L$, $N_{TE}$ and sink location $S_L$ in its buffer. The node having maximum energy will calculate minimum distance of each node within cluster called Cluster Head (CH) and broadcast $CH_{ID}$ to network nodes. It also calculates the node having minimum distance from the sink called Cluster Head Transmission (CHT) and broadcast $CHT_{ID}$ to CH. This node (CHT) will be used to transmit data toward sink if remaining CH energy is not sufficient for data transformation after computation. Once each node knows it's CH it transmits data to CH. Upon receiving data from each member node CH start computing associated sensors, missing values identification and data reduction. After this processing if CH residual energy is sufficient to transfer the data to sink it directly transfer to sink. If CH energy has not sufficient to transfer data then it handover data to CHT which will act as gateway and send the data towards sink.

*Algorithm2. Cluster Formation*
*Input:* $N_{ID}$, $N_L$, $T_S$, $N_{TE}$, $N_V$, $R_N$, $C_D$
*Output:* $CH_{ID}$, $CHT_{ID}$

*Node: Total Received Request from Nodes $R_N$*
*ClusterFormation ($N_{ID}$, $T_S$, $T_E$, $N_L$, $C_D$ )*
*{*
*NodePower=0*
*Set NodeDistance=$C_D$*
*NodeBroadcast($N_{ID}$, $N_L$, $T_S$, $T_E$, $N_V$) //within given $C_D$*
*Receive ($N_{ID}$, $N_L$, $N_V$) //Receive with $C_D$ sensor Ids*
*Compare data range $R_N$*
*For (Sensor $S_i$=1 to $S_i$=n, i++ )*
*{*
*//Compare each $S_i$ $R_N$ and $C_D$ to*
*If ($R_{Ni}$- $R_{Ni+1}$<=$R_N$ && $C_{Di}$- $C_{Di+1}$<=$C_D$)*
*{*
*JoinCluster ($R_{Ni+1}$)*

    *For( i=1 to $i_{Rn}$)*
    *{*
    *Calculate (NodePower<Maximum ($T_{Ei}$) && NodeDistance<Minimum($S_L$-$N_{Li}$))*
    *Set NodePowerID=Maximum ($E_{Si}$)*
    *Set NodeDistanceID=Minimum($S_L$-$N_{Li}$)*
    *}*
    *i ++*

*Return (NodePowerID $CH_{ID}$, NodeDistanceID $CHT_{ID}$)*

Following example along the test data explain the process of data extraction approach from a random cluster of network. Let S = {$s_1$, $s_2$, $s_3$ ...... $s_n$} be the sensors in a particular sensor network. Let the timeslot $T_S$ size equals to 5 minutes and the historical period $H_P$ is 35 minutes.

Assume the extraction process is initiated at time 08:00. At end of each time slot the nodes sensed in that timeslot will broadcast ($N_{ID}$, $T_S$, $N_E$, $N_L$, $N_V$) to neighboring nodes within given $R_N$ and $C_D$ provided by the sink to form clusters. Node having maximum energy will identify the two others nodes having minimum distance from sink called *Cluster Head Transmitter CHT* and maximum energy within cluster members called *Cluster Head* CH. Table.2 shows the detected events with in the sensor network within each timeslot. At the end of the first timeslot at (08:05), sensors *($s_2$, $s_3$, $s_4$, $s_5$)* send the messages respectively. The same process is repeated periodically for each timeslot until the end of the historical period. Table.3 shows the extracted epochs after one historical period of 35 minutes from one network cluster as an example.

Table 2: Sensor readings each timeslot

| $T_S$ | $N_{ID}$ | $N_V$ | $T_S$ | $N_{ID}$ | $N_V$ | $T_S$ | $N_{ID}$ | $N_V$ |
|---|---|---|---|---|---|---|---|---|
| 1 | $S_2$ | 1 | 2 | $S_4$ | 6 | 5 | $S_2$ | 4 |
| 1 | $S_3$ | 3 | 2 | $S_5$ | 5 | 5 | $S_3$ | 3 |
| 1 | $S_4$ | 4 | 3 | $S_4$ | 4 | 6 | $S_3$ | 7 |
| 1 | $S_5$ | 3 | 3 | $S_5$ | 5 | 6 | $S_5$ | 6 |
| 2 | $S_1$ | 2 | 4 | $S_3$ | 1 | 7 | $S_3$ | 4 |
| 2 | $S_2$ | 3 | 4 | $S_4$ | 3 | 7 | $S_4$ | 2 |
| 2 | $S_3$ | 2 | 4 | $S_5$ | 4 | | | |

After the historical period CH start processing to identify the frequent sensors to identify association rules among sensors and the estimation of missing values.

Table 3: Data stored at CH

| $T_S$ | ($N_{ID}$, $N_V$) |
|---|---|
| 1 | ($S_2S_3S_4S_5$, 1343) |
| 2 | ($S_1S_2S_3S_4S_5$, 23265) |
| 3 | ($S_4S_5$, 45) |
| 4 | ($S_3S_4S_5$, 134) |
| 5 | ($S_2S_3$, 43) |
| 6 | ($S_3S_5$, 76) |
| 7 | ($S_3S_4$, 42) |

Algorithm.3 is used to estimate missing sensor value which is identified by finding the frequent sensors where a sensor has its support *S* is higher than given *S* reported in set of epoch. If the numbers of sensors are denoted by *n* then the maximum number of possibly existing frequent sensors are:

$$\max\_num\_freq\_sensor = \sum_{i=1}^{n} \binom{n}{i}$$

Table 4.a. shows the individual frequency of each sensor reported data in a given historical period. Suppose given support S>3, Table 4.b. shows the frequent sensors at level-1 where support is higher than given support *S*. In the next step we generated the level-2 frequent sensor from level-1 by calculating the *S* in set of epoch where both appear together in same timeslot as shown in Table.4.c. Table.4.d shows level-2 frequent sensors having *S>3*.

Table: 4.a Sensor NF

| $N_{ID}$ | $N_F$ |
|---|---|
| $S_1$ | 1 |
| $S_2$ | 3 |
| $S_3$ | 6 |
| $S_4$ | 5 |
| $S_5$ | 5 |

Table: 4.b Sensor NF>3

| $N_{ID}$ | $N_F$ |
|---|---|
| $S_3$ | 6 |
| $S_4$ | 5 |
| $S_5$ | 5 |

Table: 4.c Sensor level-2 NF

| $N_{ID}$ | $N_F$ |
|---|---|
| $S_3S_4$ | 3 |
| $S_3S_5$ | 4 |
| $S_4S_5$ | 4 |

Table: 4.d Sensor level-2 NF>3

| $N_{ID}$ | $N_F$ |
|---|---|
| $S_3S_5$ | 4 |
| $S_4S_5$ | 4 |

After identification of frequent sensors at level-2 they are used to find association rules. Association rules are in the form $(S_1 \rightarrow S_2)$. The frequency of rule $(S_1 \rightarrow S_2)$ is the frequency of the $(S_1 \cup S_2)$. The *confidence* of the rule is defined as*:*

*Conf $(S_1 \rightarrow S_2)$ = Freq $(S_1 \cup S_2, DS)$ / Freq $(S_1, DS)$*

Following rules are identified from frequent sensor $(s_3s_5)$ and $(s_4s_5)$. The confidence value is set to 60 %.

*Conf $(S_3 \rightarrow S_5)$ = Freq$(S_3 \cup S_5, DS)$/ Freq$(S_3, DS)$ =66%*
*Conf $(S_5 \rightarrow S_3)$ = Freq$(S_5 \cup S_3, DS)$/ Freq$(S_5, DS)$ =80%*
*Conf $(S_4 \rightarrow S_5)$ = Freq$(S_4 \cup S_5, DS)$/ Freq$(S_4, DS)$ =80%*
*Conf $(S_5 \rightarrow S_4)$ = Freq$(S_5 \cup S_4, DS)$/ Freq$(S_5, DS)$ =80%*

The rules are used to estimate the values of sensors in those timeslots where the associated sensors have not reported data. By analyzing the values of associated sensors we can identify the upper and lower bound value range reported in each historical period. Table.5 shows the value of associated sensors pairs identified from Table 3.

Table 5: Lower and upper bound sets

| $N_{ID}$ | Pairing Values |
|---|---|
| $S_3S_5$ | (3-3), (2-5), (1-4), (7-6) |
| $S_4S_5$ | (4-3), (6-5), (4-5), (3-4) |

The missing pair of associated sensor is identified for each time slot. In Table.3 it can be observed that in timeslot 3, 5, 6 and 7 the associated sensor pair is missing. The value of missing pair is added by using the Average Round (AR) approach. For example in timeslot 3 value of $s_3$ is missing. It can be estimated from values available in Table.5. Now initial set of epoch contains values from missing pair of associated sensor shown in Table.6.

Table.6: Final DS along estimated values

| $T_S$ | $(N_{ID}, N_V)$ |
|---|---|
| 1 | $(S_2S_3S_4S_5, 1343)$ |
| 2 | $(S_1S_2S_3S_4S_5, 23265)$ |
| 3 | $(S_3S_4S_5, 445)$ |
| 4 | $(S_3S_4S_5, 134)$ |
| 5 | $(S_2S_3S_5, 434)$ |
| 6 | $(S_3S_4S_5, 756)$ |
| 7 | $(S_3S_4S_5, 425)$ |

---

*Algorithm3. Missing Value Estimation*
***Input:*** *Epoch contains Missing Values*
***Output:*** *Epoch contains Estimated Values*

***CH Node:***
*//Traverse Readings from 1 to Max Reading Size for frequent sensor readings*
*Set CountSensor =Count $(S_i)$*
    *For(i=1 to 2, i++)*
    *{*
    *call Frequent(frequentepoch)*
    *}*
    *call Frequent()*
    *{*
    *For (i=1 to N= $H_P / T_S$; i ++)*
    *{*
    *Set CountSensor =Count $(S_i)$*
    *If(CountSensor>=S) //where S is given Support value*
    *Set Frequency [SFi,]=Si[ CountSensor]*
    *Else*
    *Do Nothing*
    *Return frequentepoch*
    *}*
*\\Traverse epoch for frequent sensor's readings to estimate missing value in each window slot*
    *For ($[S_i]$ to $[S_n]$)*
    *{*
    *If( SFi>=S)*
    *{*
    *HighBound$[S_i]$= Get (Max(Epoch))*
    *LowBound$[S_i]$= Get (Min(Epoch))*
    *Estimated $[S_i]$=Avg(HighBound$[S_i]$, LowBound$[S_i]$)*
    *}*
*//Traverse Epoch for Find Missing $S_i$*
    *For (i=1 to HT / WS; i ++)*
    *{*
    *If (Traverse Epoch=Found)*
        *Set Postion$S_i$= Estimated $[S_i]$*
        *Else*
        *Do nothing}}*
*return ESTEpoch*

Table.7: DS before reduction

| $T_S$ | $(N_{ID}, N_V)$ |
|---|---|
| 1 | $(S_2S_3S_5S_4, 1334)$ |
| 2 | $(S_1S_3S_2S_4S_5, 22365)$ |
| 3 | $(S_3S_4S_5, 445)$ |
| 4 | $(S_3S_4S_5, 134)$ |
| 5 | $(S_3S_5S_2, 344)$ |
| 6 | $(S_4S_5S_3, 567)$ |
| 7 | $(S_4S_3S_5, 245)$ |

Table.8: DS after reduction

| $T_S$ | $(N_{ID}, N_V)$ |
|---|---|
| 1 | $(S_2S_3S_5S_4, 134)$ |
| 2 | $(S_1S_3S_2S_4S_5, 2365)$ |
| 3 | $(S_3S_4S_5, 45)$ |
| 4 | $(S_3S_4S_5, 134)$ |
| 5 | $(S_3S_5S_2, 34)$ |
| 6 | $(S_4S_5S_3, 567)$ |
| 7 | $(S_4S_3S_5, 245)$ |

By applying estimation process the data size *DS* is increased to reduce *DS* we applied reduction process as shown in Algorithm 4. This process first sorts the reported values as shown in Table.7. It can be observed that it contains same reported value in same timeslot within cluster. Data reduction process identifies these values and removes it from the *DS* by using *Right Trim rule*. After data reduction process the *DS* as shown in Table.8 transmitted to sink.

---

***Algorithm4. Data Reduction***
*Input: ESTEpoch*
*Output: FinalEpoch*
*CH Node:*
*//Traverse Epoch to find duplicate values from different sensor IDs in same WS*
*ApplyReduction(ESTEpoch)*
*{*
*For ( i=1 to HT/WS of ESTEpoch; i++)*
 *{*
  *While (NIDBuffer==Finish)*
  *{*
   *Traverse ([NID$_i$],[Value$_i$])*
   *If ([NID$_i$],[Value$_i$])== ([NID$_{i+1}$],[Value$_{i+1}$])*
   *Set Position=Position([NID$_i$],[])*
   *Else*
   *Next Match NID Value with Initial Value*
  *}*
*Return FinalEpoch }                    \\Send FinalEpoch to SINK*

---

This same process is executed on each cluster within network and each CH computes these values before sending *DS* towards sink. After this computation it may be possible that CH energy will not remain enough to transmit, so it can transmit *DS* to CHT, because it has the minimum distance from sink or neighboring cluster head and having sufficient energy to transmit towards sink.

Sink will receive the *DS* along estimated values and identified association rules after each historical period. Before the start of next historical period sink will broadcast these rules along other parameters for clustering formation process. After each round new rules will be evaluated on sink from historical datasets for efficient network clustering.

## 4. Experimental Results

We evaluated the performance of DDE algorithm using NS2 simulator. All experiments are based on 2.2 GHz computer with 2GB RAM and Windows XP operating system. In the network of 300 nodes, all nodes are homogenous and deployed randomly. We compared the DDE with LEACH in term of network lifetime, number of cluster heads, messages delivered, data size and number of rounds.

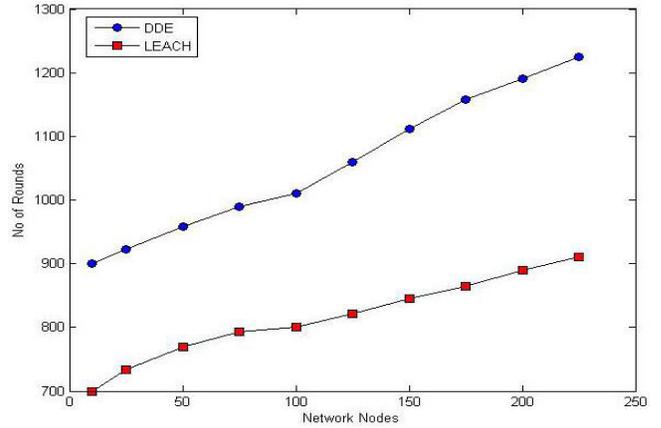
Fig.2 No of Rounds

Figure.2 shows the impact of number of rounds on network lifetime; DDE shows the good behavior if the networks size grows, whereas LEACH has less impact on network lifetime as compared to DDE.

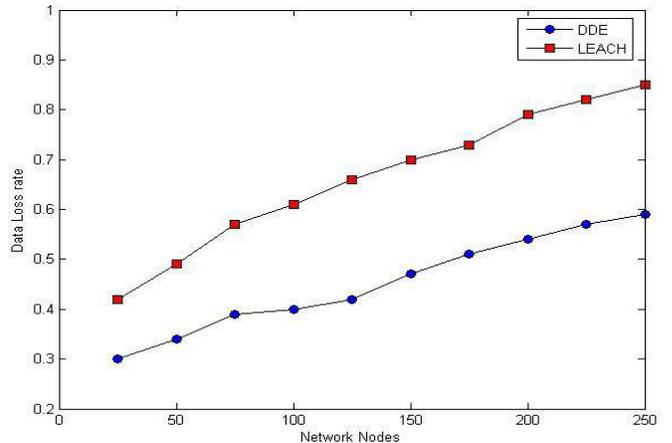
Fig.3 Data loss rate

Due to data estimation algorithm the data loss rate is also less in DDE as shown in Figure.3. The number of messages broadcast is higher in LEACH which results into more energy consumption. When the data loss is low it also consume energy but in DDE data loss is handled after data extraction step as compared to LEACH, DDE consume less energy and message broadcast during data extraction. The energy consumption and message broadcast during data extraction process is also improved as shown in Fig.4 and Fig.5.

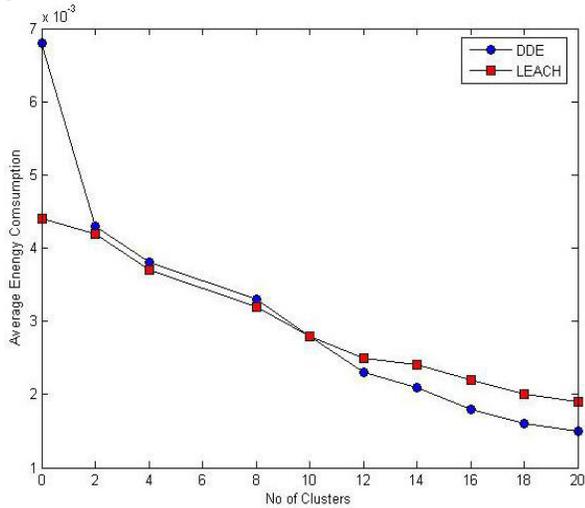
Fig.4 Avg. energy consumption

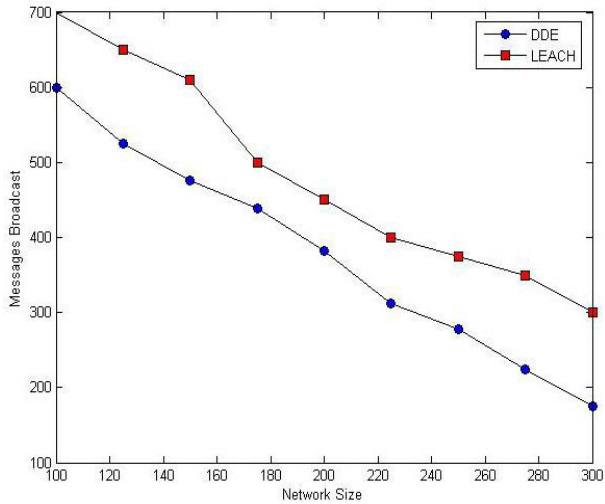
Fig.5 Messages broadcast

LEACH uses the random cluster head scheme in each network block so the numbers of force cluster heads are also increased whereas DDE uses data value range, sink distance and residual energy to create cluster and cluster heads. When number of rounds reaches more than 500 it nearly close to DDE because the numbers of still alive nodes and their residual energy remain less within network as shown in Fig.6 but during the initial rounds DDE has less no of force cluster heads.

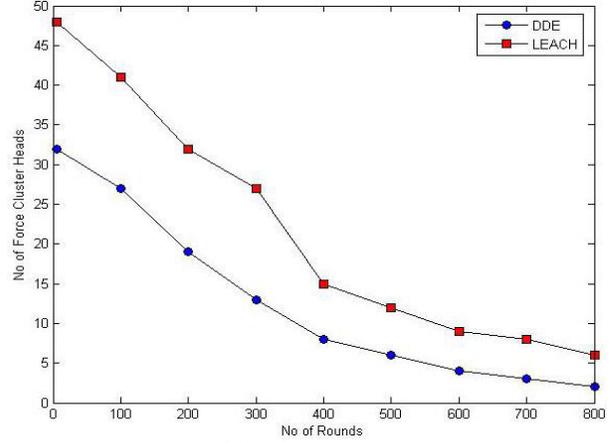
Fig.6 No of force cluster heads

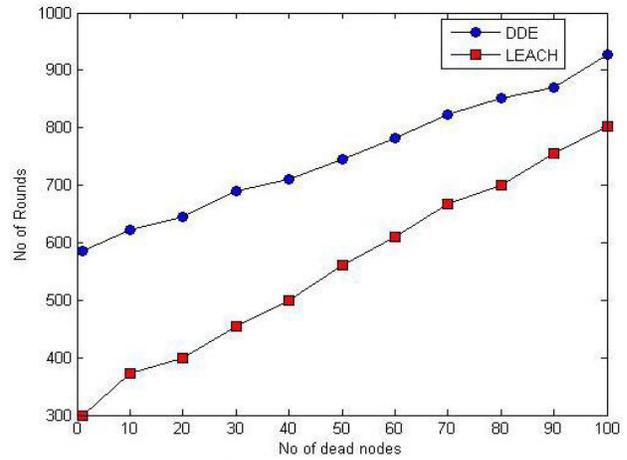
Fig.7 No of dead nodes

Sensor nodes are energy-constrained, so the network's lifetime is important for SNs application. When the number of dead node increases, the network cannot make more contributions. Thus, the network lifetime should be defined as the time when enough nodes are still alive to keep the network operational. As shown in Fig.7 LEACH has more no of dead nodes in initial rounds whereas DDE retains maximum number of nodes alive. If we compared for equal number of rounds in LEACH 100 nodes are dead in 802 rounds whereas in DDE same number of nodes are dead after 928 rounds.

## 5. Conclusion

In this paper, we have introduced a new Distributed Data Extraction (DDE) approach which consists of rule based cluster formation and identification of correlated sensor. DDE captures the temporal and data relation between the sensors by using association rule mining. The rules identified by DDE are also used to estimate the value of missing sensor within in cluster. In subsequent round these rules are used in cluster formation process where correlated sensors join the same cluster. Results show the DDE outperforms LEACH by significant margin particularly for network life time. DDE maximize the network lifetime by reducing the number of broadcast messages, energy consumption, number of dead nodes, forced cluster heads and data loss rate and maximize the number of rounds during data extraction process.

As future work, we are going to mine the extracted data on central side (SINK) to analyze the behavior of entire sensor network. By applying mining techniques at sink we can find global patterns that can be used for different purpose such as predicting the future sources of events and faulty node identification. The ongoing task of this research work is the building of adaptive data mining framework for sensor network applications.

**Acknowledgments**

This work was supported in part by the Joint Funds of NSFC-Microsoft Research Asia under Grant No. 60933012, the Specialized Research Fund for the Doctoral Program of Higher Education under Grant No.20110142110062 and International S&T Cooperation Program of Hubei Province under Grant No. 2010BFA008.